\documentclass{camera}
\usepackage{graphicx}  

\begin{document}

%
\title{MCNP6 simulation of light and medium nuclei
fragmentation 
at intermediate energies}

%
\author{Stepan G. Mashnik$^1$ \and  Leslie M. Kerby$^{1,2}$}

%
\organization{
$^1$Los Alamos National Laboratory, Los Alamos, NM 87545, USA 
$^2$University of Idaho, Moscow, Idaho 83844-4264, USA 
}

\maketitle

\begin{abstract}
Fragmentation reactions induced on light and medium 
nuclei
by protons and light nuclei of energies 
around 1 GeV/nucleon and below  are studied with 
the 
Los Alamos 
transport code MCNP6 and with its 
CEM03.03 and 
LAQGSM03.03 event generators.
CEM and 
LAQGSM assume that intermediate-energy fragmentation reactions on light 
nuclei occur generally in two stages. The first stage is the intranuclear 
cascade (INC), followed by the second, Fermi breakup disintegration of 
light excited residual nuclei produced after the INC. 
 CEM and LAQGSM 
account also for coalescence of light fragments (complex particles) up 
to $^4$He from energetic nucleons emitted during INC. We investigate the 
validity and performance of MCNP6, CEM, and LAQGSM in simulating fragmentation 
reactions at intermediate energies and discuss possible ways of further 
improving these codes.

\end{abstract}

%
\section{Introduction and theoretical background}

Fragmentation reactions induced by protons and light nuclei of energies 
around 1 GeV/nucleon and below on light 
target nuclei are involved in 
different applications, like cosmic-ray-induced single event upsets 
(SEU's), radiation protection, and cancer therapy with proton and ion 
beams, 
among others.
It is impossible to measure all nuclear 
data needed for such applications; therefore, Monte Carlo transport 
codes are usually used to simulate impacts associated with fragmentation 
reactions. It is important that available transport codes simulate such 
reactions as well as possible.

The Los Alamos Monte Carlo transport code MCNP6 
\cite{7}
uses by default the latest 
version of the cascade-exciton model (CEM) as incorporated in its event 
generator CEM03.03 
to simulate fragmentation of light nuclei at 
intermediate energies for reactions induced by nucleons, pions, and 
photons, and the Los Alamos version of the quark-gluon string model 
(LAQGSM) as implemented in the code LAQGSM03.03 
(see \cite{9} and references therein)
to simulate 
fragmentation reactions induced by nuclei and by particles at 
energies above $\sim 3.5$ GeV, up to about 1 TeV/nucleon.

Generally, both CEM and LAQGSM assume that nuclear reactions occur in three 
stages. The first stage is the IntraNuclear Cascade (INC), completely 
different in CEM and LAQGSM, in which primary particles can be re-scattered 
and produce secondary particles several times prior to absorption by, or escape 
from the nucleus. When the cascade stage of a reaction is complete, CEM and LAQGSM
use the coalescence model to ``create" high-energy d, t, $^3$He, and $^4$He via 
final-state interactions among emitted cascade nucleons, already outside of 
the target. The subsequent relaxation of the nuclear excitation is treated in 
terms of an improved version of the modified exciton model of preequilibrium 
decay followed by the equilibrium evaporation/fission stage of the reaction. 
But if the residual nuclei after the INC have atomic numbers with $A < 13$, CEM 
and
LAQGSM use the Fermi breakup model to calculate their further disintegration 
instead of using the preequilibrium and evaporation/fission models.
Thus, for targets with  $A < 13$, 
reactions are assumed to
occur only in two stages.

The ``standard'' version of CEM and LAQGSM 
account for possible multiple
emission of only n, p, d, t, $^3$He, and $^4$He during the preequilibrium
stage of reactions (see Ref. \cite{9}). Their latest, ``F'', version
(see Refs. \cite{NIMA2014, NIMB2015, Kerby-thesis}) considers a possibility
of preequilibrium emission of light fragments (LF) heavier than $^4$He,
up to $^{28}$Mg. It also simulates coalescence of LF heavier than 
$^4$He, up to $A = 7$, in CEM03.03F 
(see \cite{NIMA2014, NIMB2015, Kerby-thesis}), and up to $A = 12$, in LAQGSM03.03F
(see \cite{NUFRA2015}).

In recent years, MCNP6, with its CEM and LAQGSM event generators,
has been
extensively validated and verified (V\&V) against 
a large variety of nuclear reactions on both thin and thick targets 
(see, e.g. Refs. 
\cite{NIMA2014} - \cite{12}
 and references therein). 
In Ref. \cite{NIMA2014}, it was 
tested
specifically 
on fragmentation of light nuclei at intermediate energies.
Here, we present a few results from our recent
work \cite{NIMA2014}
and
investigate further
the performance of MCNP6, CEM, and LAQGSM in 
simulating fragmentation reactions at intermediate energies 
and discuss 
possible ways of further improving these codes.

\section{Results and conclusion}

Figs. 1, 2, 3, and 5
show examples of fragmentation reactions on light nuclei
simulated by our codes. Figs. 6 and 4 shows examples for 
medium targets, $^{48}$Ca and $^{nat}$Ag.
Many more similar results,
their discussion, and useful details
 can be found in Refs.
\cite{NIMA2014} - 
\cite{LAQGSM03.03}
and references therein.



\begin{figure}[h]
\begin{minipage}{5.0cm}
\vbox to 50mm {
\includegraphics[width=6cm]{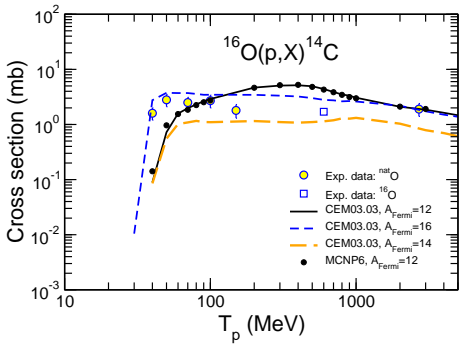}}             
\end{minipage}
\hfill
\begin{minipage}{6cm}
\vspace*{-10mm}
\caption{
Excitation function for the 
production of $^{14}$C from p + $^{16}$O 
calculated with CEM03.03 using the ``standard'' version of the Fermi 
breakup model ($A_{Fermi} = 12$) and with cut-off 
values for $A_{Fermi}$ of 16 and 14, as well as with MCNP6 using CEM03.03 
($A_{Fermi} = 12$) compared with experimental 
data, as indicated. 
Experimental data are from the T16 Lib 
compilation \cite{42} 
(see details in \cite{NIMA2014}).
}
\end{minipage}
\label{SGMfig1} 
\end{figure}

\vspace*{-5mm}
\begin{figure}[h]
\vspace*{-9mm}
\begin{minipage}{5.0cm}
\vbox to 50mm {
\includegraphics[width=6cm]{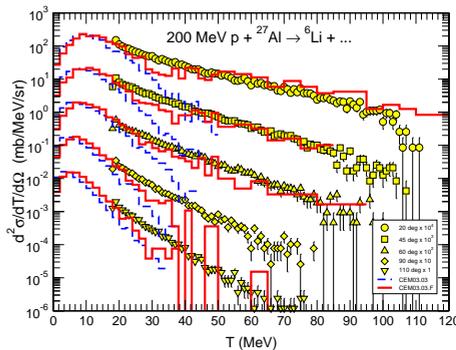}}             
\end{minipage}
\hfill
\begin{minipage}{6cm}
\vspace*{3mm}
\caption{
Comparison of experimental $^6$Li spectra at 20, 45, 60, 90, and 110 degrees 
by Machner et al. 
\cite{61}
(symbols) with calculations by the unmodified CEM03.03 (dashed 
histograms) and results 
by
CEM03.03.F 
(solid histograms), as indicated
(see more details in \cite{NIMA2014}).
}

\vspace*{5mm}
Our results indicate that MCNP6 using
CEM03.03 and LAQGSM03.03 simulates  fragmentation reactions on light
and
\end{minipage}
\label{SGMfig2} 
\end{figure}

\vspace*{-4mm}
{\noindent
 medium-light nuclei at intermediate energies well,
in a satisfactory agreement with experimental data. }
The recent ``F'' version of codes
(see Refs. \cite{NIMA2014} -- \cite{NUFRA2015})
is even better, as it allows us to describe emission
of energetic LF from practically arbitrary reactions.

However, MCNP6 is not yet ready to predict well 
heavy fragments
from reactions with heavier nuclei, with mass numbers $A \sim 100$.
Such nuclear targets are considered too light to fission
in conventional codes.
Similarly,
the fragments are too light to be produced as spallation
residues and too heavy to be produced via standard
evaporation and/or preequilibrium models, or via coalescence.
\clearpage

\begin{figure}[h]
\begin{minipage}{5.0cm}
\vbox to 50mm {
\includegraphics[width=6cm,clip]{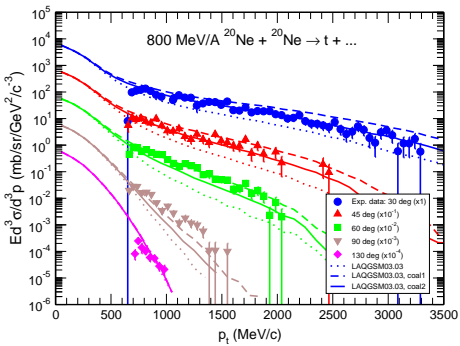}}       
\end{minipage}
\hfill
\begin{minipage}{6cm}
\vspace*{-10mm}
\caption{
Comparison  
of measured \cite{57}
t spectra at 45, 60, 90, and 130 degrees 
from 800 MeV/nucleon $^{20}$Ne +  $^{20}$Ne
with calculations by LAQGS03.03 using its ``standard'' version of the coalescence 
model ($p_0 = 0.108$ GeV/c for t and $^3$He;
dotted lines) and with modified values of $p_0$ labeled in legend as 
``coal1'' 
(dashed lines) and
``coal2'' (solid lines), as indicated in legend and 
discussed in detail in Ref. \cite{NIMA2014}.
}
\end{minipage}
\label{SGMfig3} 
\end{figure}

\begin{figure}[h]
\vspace*{-5mm}
\begin{minipage}{5.0cm}
\vbox to 50mm {
\includegraphics[width=6.5cm]{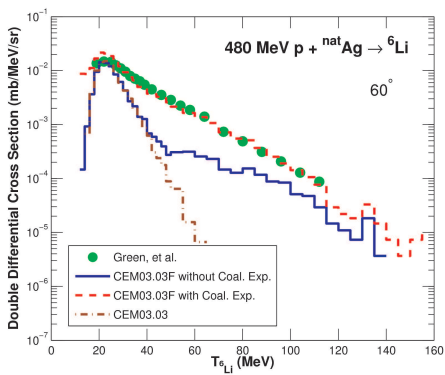}}          
\end{minipage}
\hfill
\begin{minipage}{6cm}
\vspace*{-0mm}
\caption{
Comparison of experimental data by Green {\it et al.} \cite{Green480} 
(circles) for the production of $^6$Li at an angle of 
60$^\circ$ from the reaction 480 MeV p + $^{nat}$Ag, with results 
by standard CEM03.03 (dot-dashed line)
from CEM03.03F without coalescence expansion (solid line) and 
CEM03.03F with coalescence expansion (dashed line)
(see details in \cite{Kerby-thesis, MCNP6imps}).
}

\vspace*{5mm}

One way to approach this problem would be to employ
after the INC 
stage of reactions
a fission-like 
\end{minipage}
\label{SGMfig4} 
\end{figure}

\vspace*{-4mm}
{\noindent
sequential-binary-decay model,
 like the 
code GEMINI  by Charity {\it et al.}
\cite{GEMINI}
to describe the compound nuclear
decay. }
In our case, this means separately merging
CEM and LAQGSM with GEMINI. Actually,
we already have done so more than
a decade ago, producing the ``G'' versions of 
CEM and LAQGSM we had at that time
(see, {\it e.g.}, Ref. \cite{S1G1} and references therein).

Another way to address this problem is to implement in CEM and LAQGSM
the
 Statistical Multifragmentation Model (SMM) by Botvina {\it et al.} 
\cite{SMM}.
Thus, we would
consider multifragmentation as a mode competitive
to evaporation of particles and light fragments, when the
excitation energy $E^*$ of a compound nucleus produced after the
preequilibrium stage of a reaction is above 
a certain value, $E^*_{tr}$, e.g., $E^*_{tr} = 2 \times A$ MeV, as we
did in the ``S" versions of CEM03.01 and LAQGSM03.01
(see, e.g., Ref. \cite{S1G1} and references therein). 

As of today, neither the ``S'' nor the ``G'' versions of CEM and LAQGSM have
been implemented in MCNP6. We plan to incorporate them in our event
generators used by MCNP6 after we tune several parameters in SMM 

\begin{figure}[h]
\begin{minipage}{5.0cm}
\vbox to 55mm {
\includegraphics[width=6cm,clip]{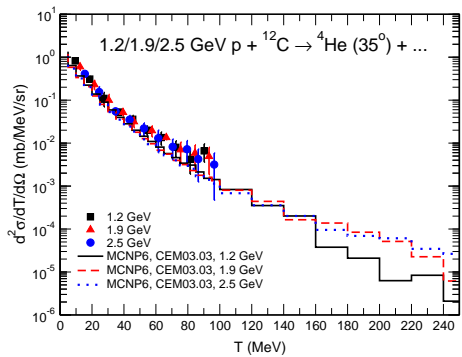}}   
\end{minipage}
\hfill
\begin{minipage}{6cm}
\vspace*{-5.5mm}
\caption{
$^4$He spectra (at 35$^\circ$) from 1.2/1.9/2.5 GeV p + $^{12}$C 
measured by M. Fidelus of the PISA collaboration 
\cite{83}
(symbols) with calculations by 
MCNP6 using CEM03.03 (see details in \cite{Kerby-thesis}).
}

\vspace*{5mm}
and GEMINI, 
that are essential in chosing the excitation energy
 (or temperature) of nuclei
when reaction mechanisms change from
 ``usual evaporation'', to binary decays described by GEMINI,  and/or

\label{SGMfig5} 
\end{minipage}
\end{figure}

\vspace*{-4.7mm}
{\noindent
 to multifragmentation simulated with SMM
(see 
details 
in \cite{MCNP6imps}).
}

\begin{figure}[h]
\begin{minipage}{5.0cm}
\vbox to 65mm {
\includegraphics[width=7cm,clip]{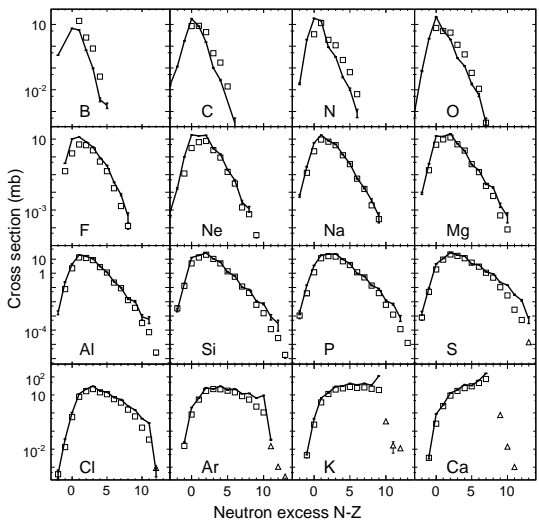}}   
\end{minipage}
\hfill
\begin{minipage}{5cm}
\vspace*{1mm}
\caption{
Measured cross sections for $^{48}$Ca fragmentation on $^9$Be
at 140 MeV/nucleon 
\cite{MockoPhD}
compared with LAQGSM03.03 predictions (see details in \cite{LAQGSM03.03}). 
}

\vspace*{5mm}
This study was carried out under the auspices of the National Nuclear 
Security Administration of the U.S. Department of Energy at Los Alamos 
National Laboratory under Contract No. DE-AC52-06NA25396.
This work is supported in part (for L.M.K)  by the M. Hildred

\label{SGMfig6} 
\end{minipage}
\end{figure}

\vspace*{-4mm}
{\noindent
 Blewett 
Fellowship of the American Physical Society, 
www.aps.org.
}

%

\end{document}